\documentclass[11pt,a4paper]{article}
\pdfoutput=1

\usepackage{graphicx}
\usepackage{color}
\usepackage[colorlinks=true, urlcolor=blue, linkcolor=blue, 
citecolor=blue]{hyperref}
\usepackage{cite}

\newcommand{\beq}{\begin{equation}}
\newcommand{\eeq}{\end{equation}}
\newcommand{\beqa}{\begin{eqnarray}}
\newcommand{\eeqa}{\end{eqnarray}}
\newcommand{\beqar}{\begin{eqnarray*}}
\newcommand{\eeqar}{\end{eqnarray*}}

\begin{tiny}

\end{tiny} 

\begin{document}
\thispagestyle{empty}
$\,$

\vspace{30pt}

\begin{center}

\textbf{\Large High energy muons in extensive air showers}

\vspace{50pt}
C.~G\'amez\footnote{Now at CIEMAT, Avenida Complutense 22, E-28040 Madrid, Spain}, 
M.~Guti\'errez, J.S.~Mart\'\i nez, M.~Masip 
\vspace{16pt}

\textit{CAFPE and Departamento de F{\'\i}sica Te\'orica y del Cosmos}\\
\textit{Universidad de Granada, E-18071 Granada, Spain}\\
\vspace{16pt}

\texttt{car,mgg,jsilverio,masip@ugr.es}

\end{center}

\vspace{30pt}

\date{\today}

\begin{abstract}
The production of very high energy muons inside an extensive air shower is 
observable at $\nu$ telescopes and 
sensitive to the composition of the primary cosmic ray. 
Here we discuss five different sources of these muons: pion and kaon
decays; charmed hadron decays; rare decays of unflavored mesons;
photon conversion into a muon pair; and photon conversion into
a $J/\psi$ vector meson decaying into muons. 
We solve the cascade equations for a $10^{10.5}$ GeV proton primary 
and find that unflavored mesons and gamma conversions 
are the two main sources of $E\ge 10^{8.5}$ GeV muons, while charm decays 
dominate at $10^{5.5}\,{\rm GeV}< E< 10^{8.5}\,{\rm GeV}$. 
In inclined events one of these muons
may deposite a large fraction of its energy near the surface, implying 
fluctuations in the longitudinal profile of the shower and in the muon to electron 
count at the ground level. In particular, we show that 
1 out of 6 proton 
showers of $10^{10.5}$ GeV include an  $E>10^6$ GeV 
deposition within 500 g/cm$^2$, while only in 1 out of 330 showers
it is above $10^7$ GeV. We also show that the production
of high energy muons is very different in proton, iron or photon showers ({\it e.g.}, 
conversions $\gamma\to \mu^+ \mu^-$ are the main source of 
$E\ge 10^4$ GeV muons in photon showers). Finally, we use Monte Carlo
simulations to discuss the validity of our results.

\end{abstract}

\newpage

\section{Introduction}
Extensive air showers (EASs) initiated by ultrahigh energy cosmic rays (CRs)
include millions of collisions and decays of secondary particles. These showers 
can be observed with fluorescence detectors, able to measure energy 
depositions as the shower develops along the 
atmosphere, and/or surface detectors, which register energy depositions of
the particles reaching the ground.
Despite the large number of processes involved in an individual EAS, its 
dynamics can be understood within the following simplified scheme \cite{Gaisser:1990vg}.

Let us take a proton primary facing a relatively large slant depth 
({\it e.g.}, $X\approx 2000$ g/cm$^2$ from a zenith angle 
$\theta_z\approx 60^\circ$). 
The first interaction will take place high in the atmosphere, 
after the proton has crossed a hadronic interaction length 
($\lambda_{\rm int} \approx 41$ g/cm$^2$ at $E=10^{10}$ GeV).
It will result into a leading baryon
carrying around 25\% of the initial energy plus dozens of 
light mesons (mostly pions) sharing the rest of the energy. The leading baryon 
will interact again deeper into the atmosphere, but after just four collisions 
99\% of its energy will already be transferred to pions. 
High energy charged pions, in turn, may collide giving more pions of lower energy
or they may decay into leptons, $\pi^+\to \mu^+ \nu_\mu$. Decays are only favoured at 
relatively low energies, below (with a strong dependence on the altitude) 
$50$ GeV, so the production of very high energy muons 
and neutrinos is suppressed. In contrast, neutral pions of all energies decay almost 
instantly ($\pi^0\to \gamma \gamma$) giving photons that feed
the electromagnetic (EM)
component of the EAS. Photons will convert into $e^+ e^-$ pairs after 
$9X_0/7\approx 47$ g/cm$^2$, whereas electrons will radiate half their energy after a
similar depth, so the EM energy is transformed fast into 
a large number of lower-energy particles that 
define the shower maximum at $X_{\rm max}\approx 800$ g/cm$^2$.
The precise position of $X_{\rm max}$ has fluctuations 
$\Delta X_{\rm max}\approx 50$ g/cm$^2$ that depend basically on the details in the first
few collisions of the leading baryon.
Notice also that most of the energy in the EAS will be processed through 
gammas and electrons instead of muons and neutrinos: although the three 
pion species are created at a similar rate, 
high-energy charged pions tend to collide giving both charged and
neutral pions, whereas all the energy that goes into neutral pions becomes EM and has a
small return to hadrons. In inclined events, after
a depth around $2X_{\rm max}$ most gammas and electrons have been absorbed
by the atmosphere 
and the signal becomes dominated by muons, although it includes an EM {\it tail} created 
by muon radiative depositions and muon decays.

Within this simple picture,  the value of $X_{\rm max}$ or the signal at the surface
detectors  depend critically on the inelasticity (fraction of energy 
lost by the leading hadron) and the multiplicity (number of secondary hadrons 
that share that energy) in nucleon and pion collisions in the air.
Unfortunately, the study of these two observables at colliders is not easy,
as it involves a very wide range of energies and a kinematical region (ultraforward) of
difficult access.\footnote{A forward spectrometer at the LHC appears indeed as a very
promising possibility \cite{Albrow:2018kxz}.}
The uncertainty that they introduce (for example, through the appearance of
collective effects \cite{Baur:2019cpv})
could possibly explain the apparent 50\% excess 
 in the number of muons at the ground level 
recently emphasized by the Pierre Auger Observatory \cite{Aab:2014pza}.
Nevertheless, the picture provides a good description of the main features in 
an EAS. The different values of $X_{\rm max}$ 
for proton or iron primaries, for example, 
are easily understood if one sees a nucleus shower as the 
superposition of $A$ nucleon showers of energy $E/A$, each one 
with a smaller value of $X_{\rm max}$.

However, there are a few features in an EAS that are in principle observable and
require a more elaborate scheme. The production of 
$E\ge 10^5$ GeV neutrinos is
one of them: It is expected that at those energies 
pion and kaon decays become a less effective source of neutrinos
than charmed hadron decays \cite{Gondolo:1995fq,Enberg:2008te}. 
Another one are the muons of also very high energy.
The weak decays of mesons produce both muons and neutrinos, but muons 
may also appear with no neutrinos in 
the EM decays of unflavored mesons or in the interactions of high energy photons 
with atmospheric nuclei \cite{Illana:2009qv,Illana:2010gh}.
These relatively rare processes are not always included 
in Monte Carlo simulations, in particular, some rare decays of unflavored mesons
are absent in EPOS-LHC  \cite{Pierog:2013ria} and only the most recent version 
of SIBYLL \cite{Fedynitch:2018cbl} includes
the production of charmed hadrons. 
Our first objective in this work is to review and compare the different production 
channels of high energy muons inside an EAS. 

We find several phenomenological 
reasons why these most energetic muons may be interesting. First, 
they could be useful in composition studies. Obviously, an iron shower will
never contain a muon carrying a fraction of energy larger than 
$1/A=0.017$; but how frequent are muons with,
for example, a 0.1\% of the shower energy?
It turns out that such muons are 10 times more likely in a proton than in
an iron EAS. They will often appear in pairs, in the core of the
shower, always accompanied by a bundle of lower energy muons.
Their inclination when they cross a 
neutrino telescope  and/or catastrophic energy loses there would reveal the 
minimum energy of these muons \cite{Aartsen:2017upd,Bogdanov:2018sfw,Kochanov:2019yvx}.

Another phenomenological reason of interest rarely explored in the literature
(see for example \cite{Tascau:2007zz})
is their possible effect on the
longitudinal development of inclined EASs. It has been shown \cite{Canal:2016bvv} 
that the ratio $r_{\mu e}$ 
of the muon to EM signals at the ground level (number of muons over total EM 
energy at the surface detectors) is strongly 
correlated with the position of the shower maximum,
and that the correlation
seems to be independent from the hadronic model used in the simulation.
Since the fluctuations between two showers from identical primaries are 
basically caused by the initial hadronic processes, 
the great stability in the $r_{\mu e}$--$X_{\rm max}$ correlation
 (due only to collisions and decays
{\it after} the shower maximum) is not surprising.
In a search for possible heavy quark effects in EASs, the analysis with the
code AIRES \cite{AIRES} in \cite{Canal:2016bvv} finds that
sometimes a very energetic muon created in $D$ or $B$ decays 
introduces anomalies in that ratio. Indeed, an  $E>10^6$ GeV  deposition  near
the ground (when most of the shower energy has already been absorbed) in
an inclined ($\theta_z\ge 60^\circ$) event could
change substantially the signal at the surface detectors. The programmed 
upgrade at AUGER \cite{Aab:2016vlz}
may provide a more precise separation of the muon and EM signals, so it seems
interesting an estimate of how frequent such muon energy depositions are and what
their origin (in addition to heavy quark decay) may be. This is precisely our second
objective in this work.

The plan of this article is as follows. In the next section we review
the different processes that may produce high energy muons 
inside an EAS. In section 3 we solve the cascade equations for 
the average $10^{10.5}$ GeV proton, iron or gamma shower 
and we deduce the spectrum of muons reaching the ground from large zenith 
inclinations. In section 4 we parametrize the three main radiative processes
experienced by a muon  in the air (bremsstrahlung, pair production and photonuclear 
collisions) and we estimate the probability for a large energy deposition near the 
surface. Finally, in section 5 we discuss the validity of our method by comparing
with Monte Carlo simulations and we conclude 
in section 6.

\section{Muon production channels}

\noindent {\it (i) Conventional muons from pion and kaon decays.} \\
We have obtained a fit\footnote{The details  
about these fits will be presented elsewhere.}
for the yields $f_{hh'}(x,E)$ of hadrons $h'=p,n,\bar p,\bar n,\pi^\pm,K^\pm,K_L$ produced 
in the collisions of nucleons, pions and kaons of energy $E$ with an average air nucleus
($x=E_{h'}/E$).
In particular, the four lowest moments provided by our fits, 
\beq
Z_{hh'}(n,E)=\int_0^1 {\rm d}x\; x^n\; f_{hh'}(x,E)\,,
\eeq
match the ones derived from $5\times 10^4$ collisions simulated with EPOS-LHC 
(using the crmc package \cite{crmc}) at different energies. 
\begin{figure}[!t]
\begin{center}
\includegraphics[width=0.455\linewidth]{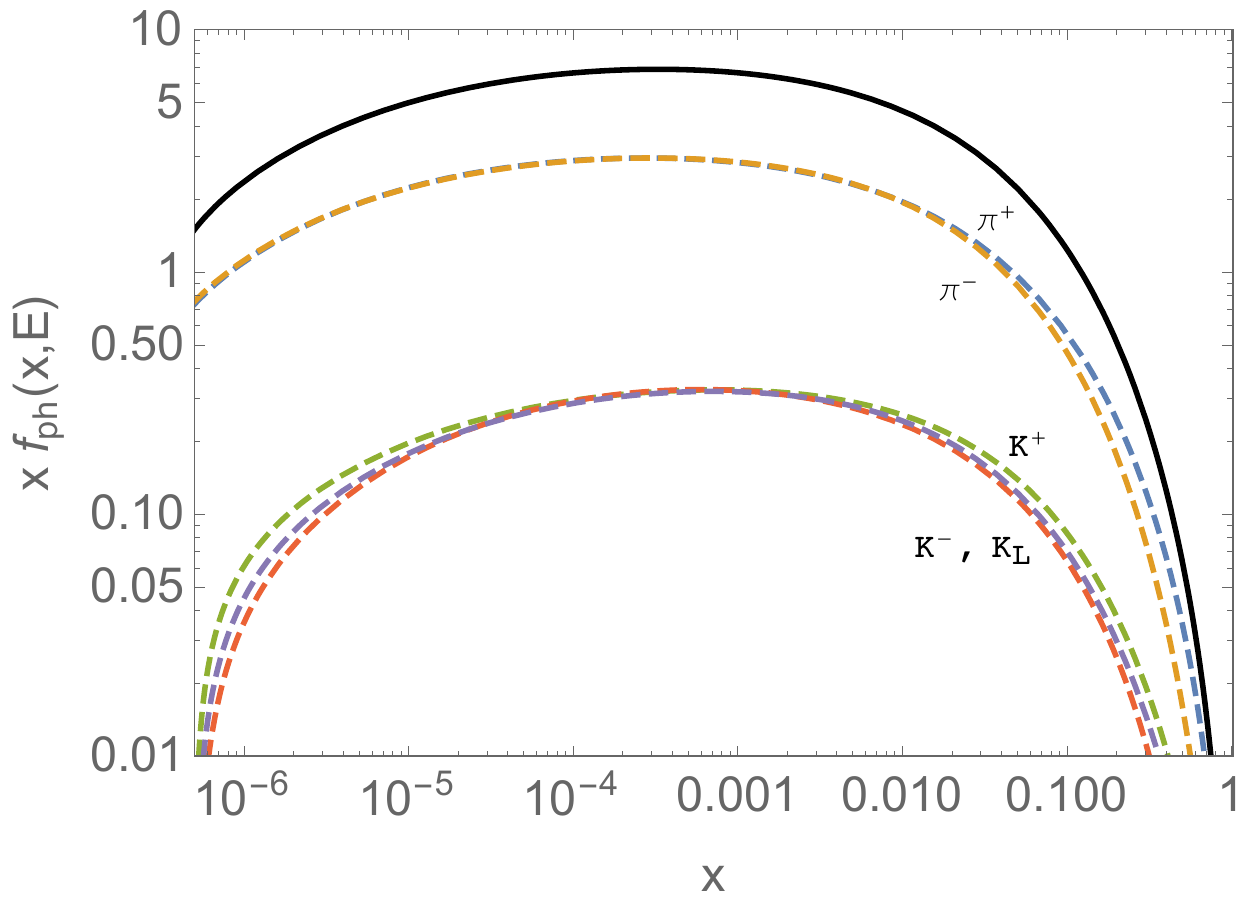}\hspace{0.8cm}
\includegraphics[width=0.41\linewidth]{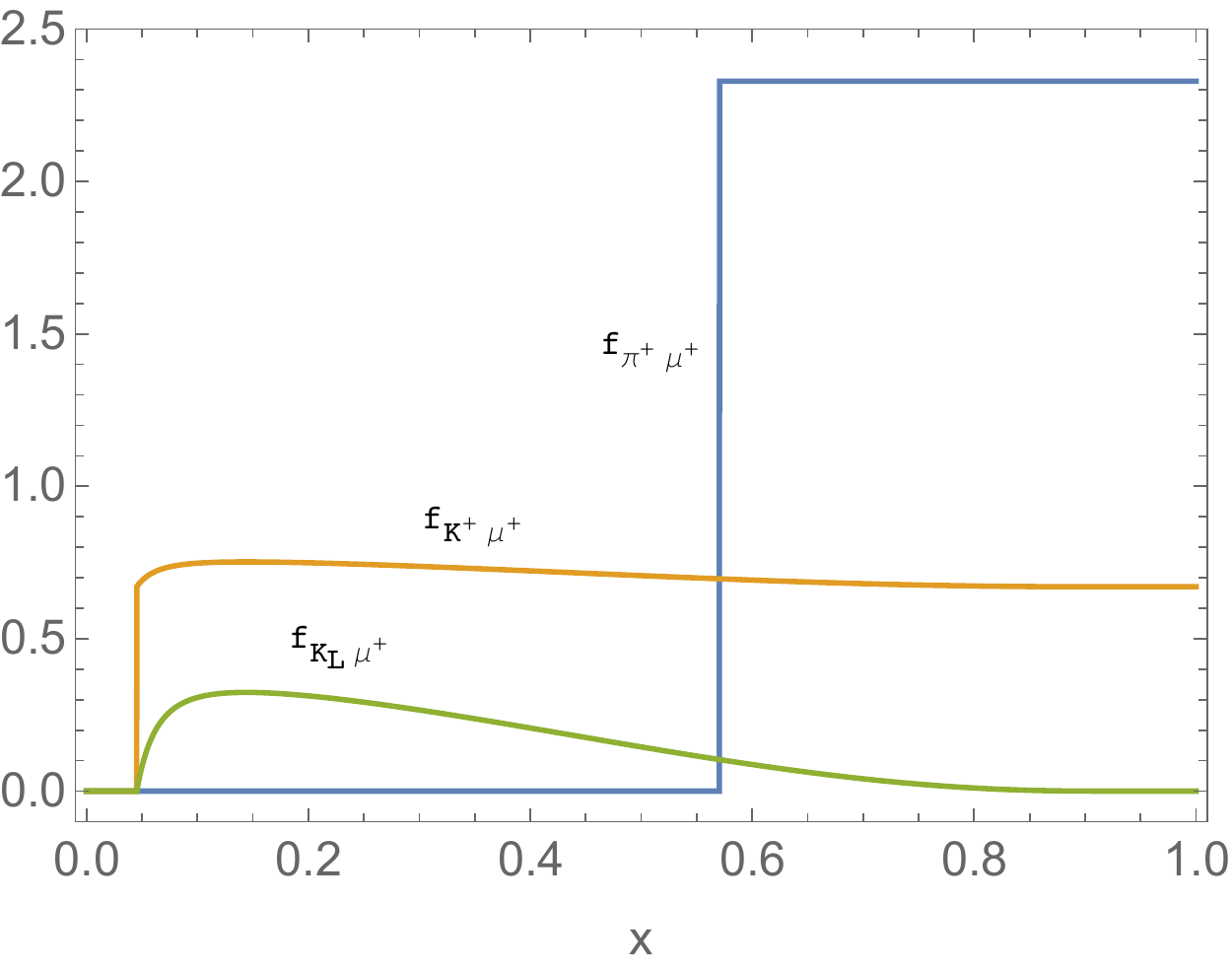}
\end{center}
\vspace{-0.5cm}
\caption{Pion and kaon yields in proton-air collisions at $10^6$ GeV obtained
with EPOS-LHC (left) and 
muon yields in pion and kaon decays (right).
}
\label{fig1}
\end{figure}
In Fig.{1}--left we show for illustration the 
yields of light mesons in proton--air collisions at $10^6$ GeV. 
Notice that the zero moment of $f_{hh'}(x,E)$ corresponds
to the total number of particles $h'$ created per collision, whereas the first moment is
the fraction of energy taken by these particles. In the example, 
the average collision produces 52.9 charged pions and 7.9 kaons that take respectively 
34.6\% and 7.1\% of the proton energy. Using 
SIBYLL 2.3C \cite{Ahn:2009wx} we obtain similar results:  51.6 charged
pions and 8.5 kaons carrying 30.7\% and 7.6\% of the proton energy, respectively. The yields 
in pion and kaon collisions are analogous.

The decay of these light mesons will produce muons \cite{Lipari:1993hd}.
In Fig.~1--right we have included the decay yields
$f_{h\mu}(x,E)$ in the ultrarelativistic limit, with the zero 
moment giving the  branching ratio into muons in these decays.

\noindent {\it (ii) Muons from charmed-hadron decays.} \\
SIBYLL 2.3C has included charmed-hadron production: in Fig.~2 we plot
for illustration the yields in proton collisions at $10^6$ GeV.  
It is remarkable that these yields incorporate at least a fraction of
the so called {\it forward} 
charm \cite{Brodsky:1980pb,Halzen:2016thi,Carvalho:2017zge}. This refers to
charm produced through a matrix element at any order that 
combines with a (spectator) valence quark of the projectile ({\it i.e.}, coalescence
in the fragmentation region) or 
charm produced in diffractive collisions (after pomeron exchange the 
diffractive mass of the projectile is large enough to give a couple of charmed
hadrons).
In both cases the collision results into
a forward charmed hadron carrying a large fraction $x$ of the collision energy.
Notice that perturbative calculations 
\cite{Gauld:2015kvh} combine amplitudes with fragmentation functions
that (according to factorization theorems) do not depend on the initial state, whereas
these codes use a fragmentation model that allows coalescence. 

\begin{figure}[!t]
\begin{center}
\includegraphics[width=0.50\linewidth]{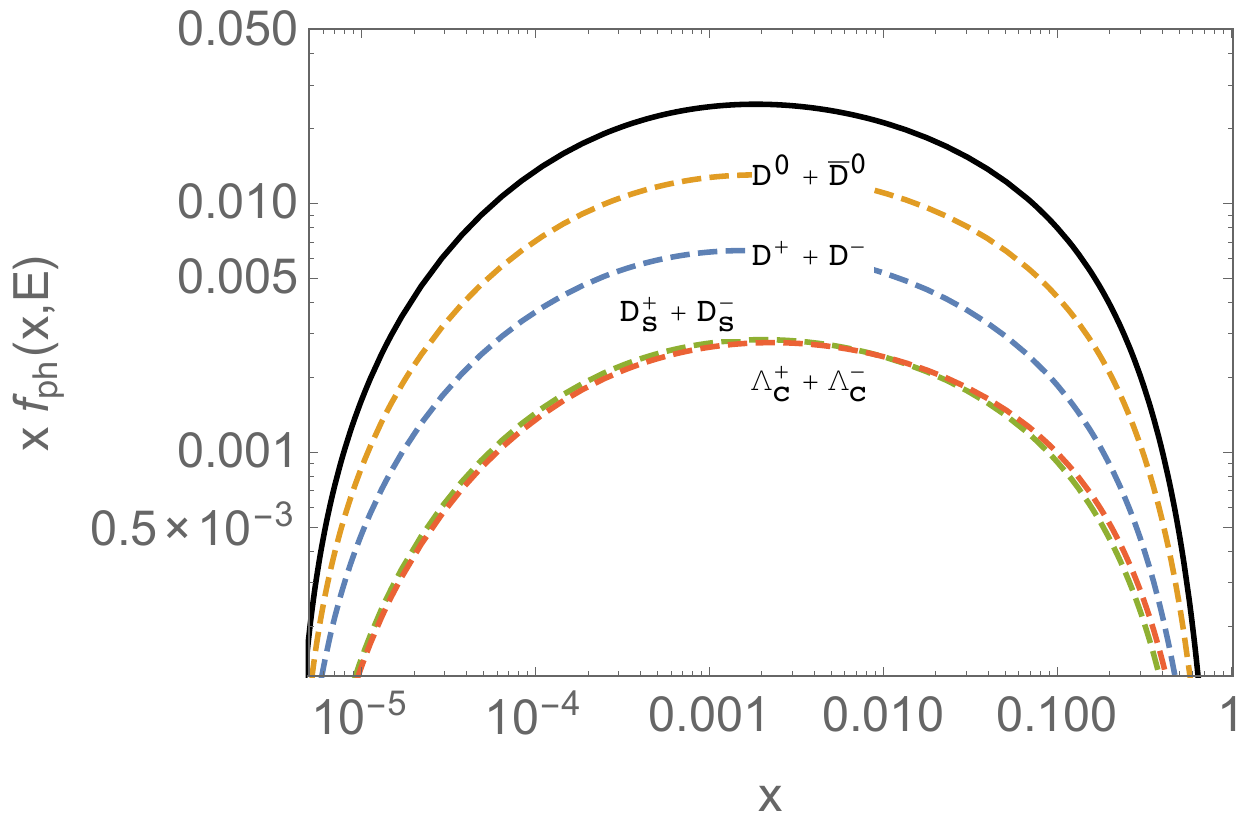}\hspace{0.8cm}
\includegraphics[width=0.417\linewidth]{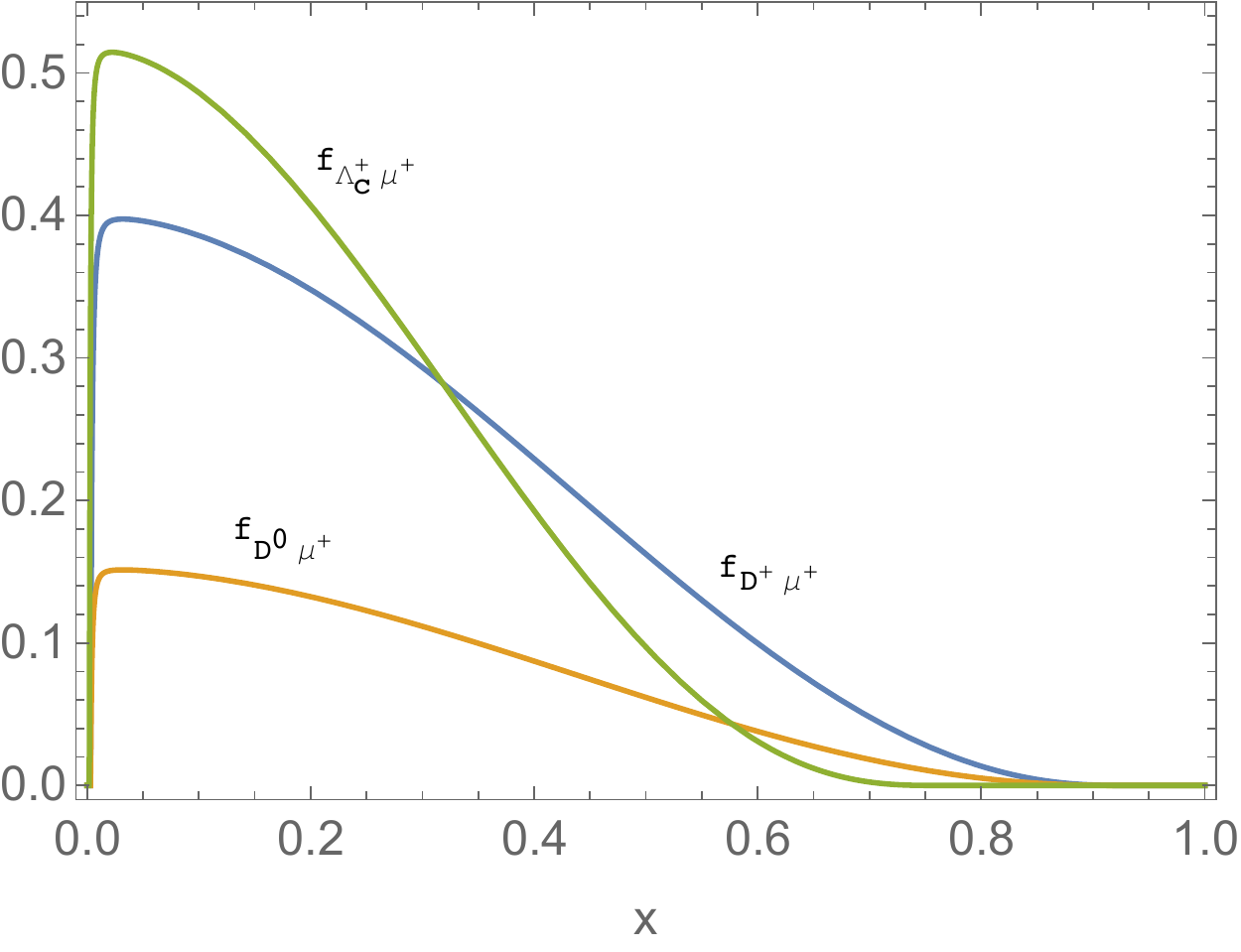}
\end{center}
\vspace{-0.5cm}
\caption{Left: Charmed-hadron yields in proton-air 
collisions at $10^6$ GeV obtained with SIBYLL 2.3C \cite{Ahn:2009wx}. Right: Muon yields in
charmed hadron decays.
}
\label{fig2}
\end{figure}

Once produced the $D$ mesons and $\Lambda_c$ baryons 
may decay giving muons (see the decay yields in Fig.~2 \cite{Illana:2010gh}). 
However, at 
$E\ge 10^6$ GeV they may also collide loosing part of their energy. The inelasticity
$K=1-\langle x\rangle$
in charmed hadron collisions with air is smaller than in pion or proton collisions.
In \cite{Barcelo:2010xp,Bueno:2011nt} $\langle x\rangle$ is estimated with
PYTHIA \cite{Sjostrand:2006za} simulating light hadron collisions and  then
replacing (after the collision but before
fragmentation) the leading up quark by a charm
quark.
The results for the fraction of 
energy carried by  the leading charmed hadron after the collision 
can be approximated by a gaussian distribution with 
$\langle x\rangle=0.56$,
versus just $\langle x\rangle=0.26$ for the leading pion in a $10^6$ GeV
pion collision. 

\noindent {\it (iii) Muons from the rare decays of unflavored mesons.} \\
The unflavored mesons $\eta$, $\rho$, $\omega$, $\eta'$ and $\phi$, with masses between
0.5 and 1 GeV, include decay channels with muon pairs. For example, 
BR$(\eta\to \mu^+\mu^-\gamma)=3.1\times10^{-4}$ or 
BR$(\phi\to \mu^+\mu^-)=2.9\times10^{-4}$ \cite{Tanabashi:2018oca}. These decay modes 
are more rare than in 
$D$-meson decays ({\it e.g.}, BR$(D^+\to \bar K^0\mu^+ \nu_\mu)=0.092$), but
this is partially compensated by the smaller mass and then the 
larger frequency of unflavored mesons 
in hadronic collisions.
Moreover, they always decay promptly, whereas most 
$D$ mesons and $\Lambda_c$ baryons of 
$E> 10^7$ GeV collide in the air and lose energy instead of decaying.

\begin{figure}[!t]
\begin{center}
\includegraphics[width=0.46\linewidth]{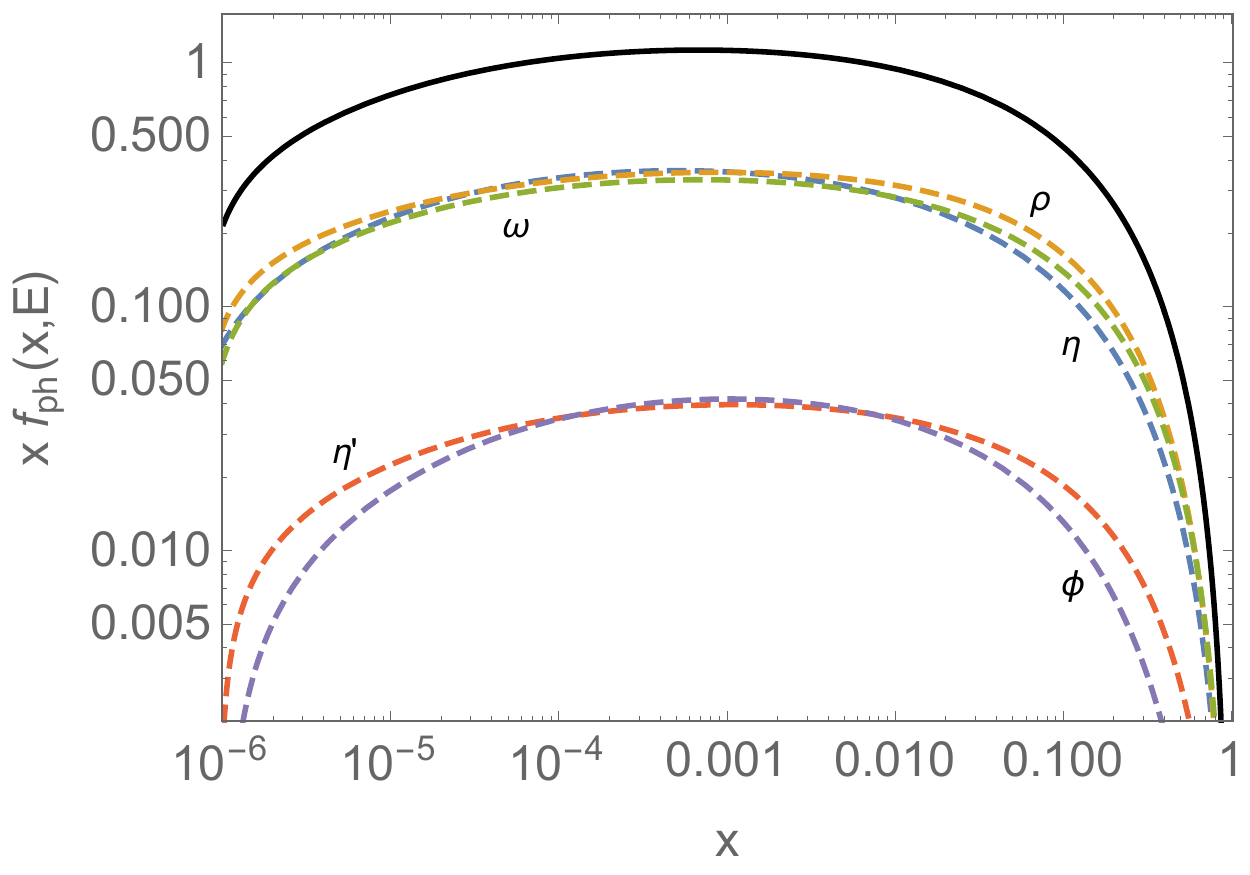}\hspace{0.8cm}
\includegraphics[width=0.44\linewidth]{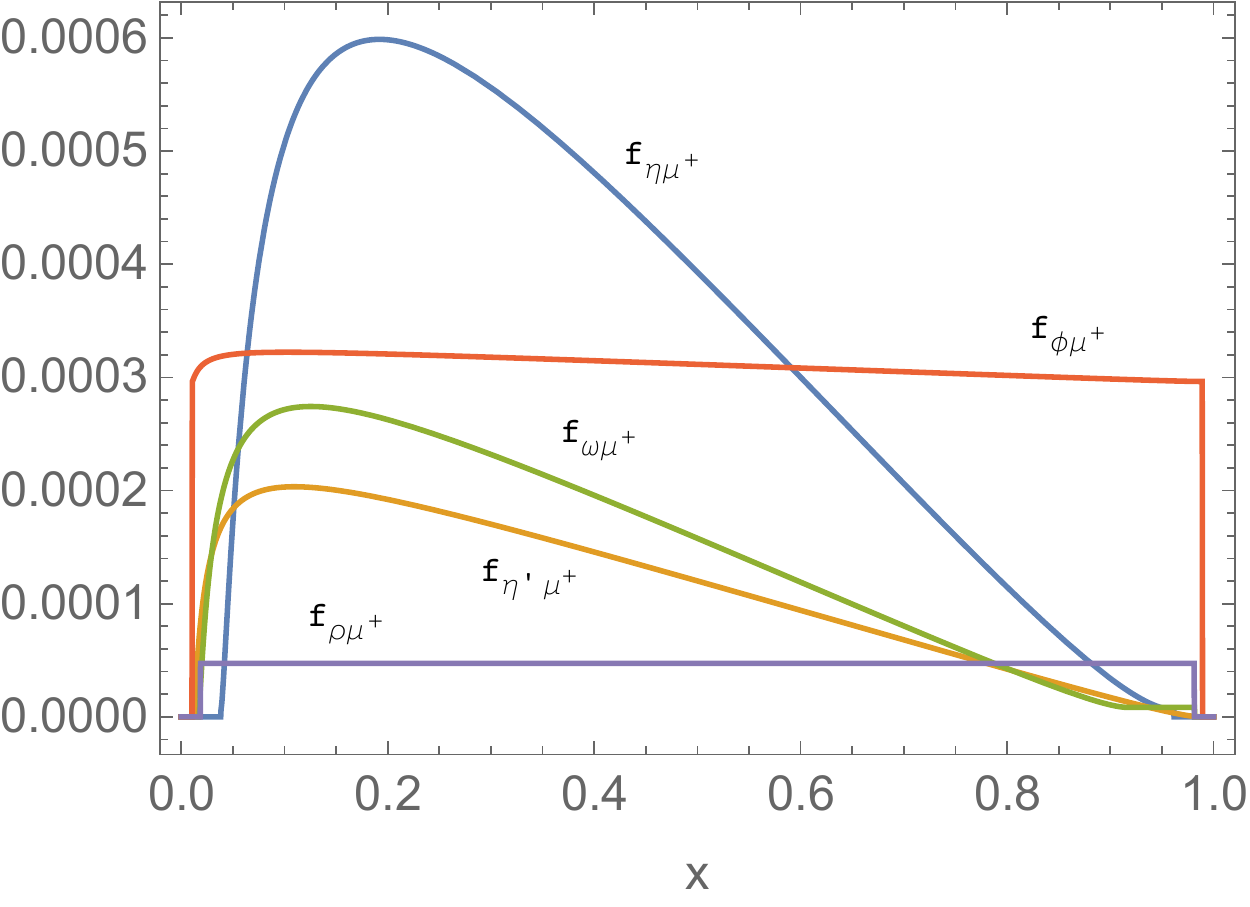}
\includegraphics[width=0.48\linewidth]{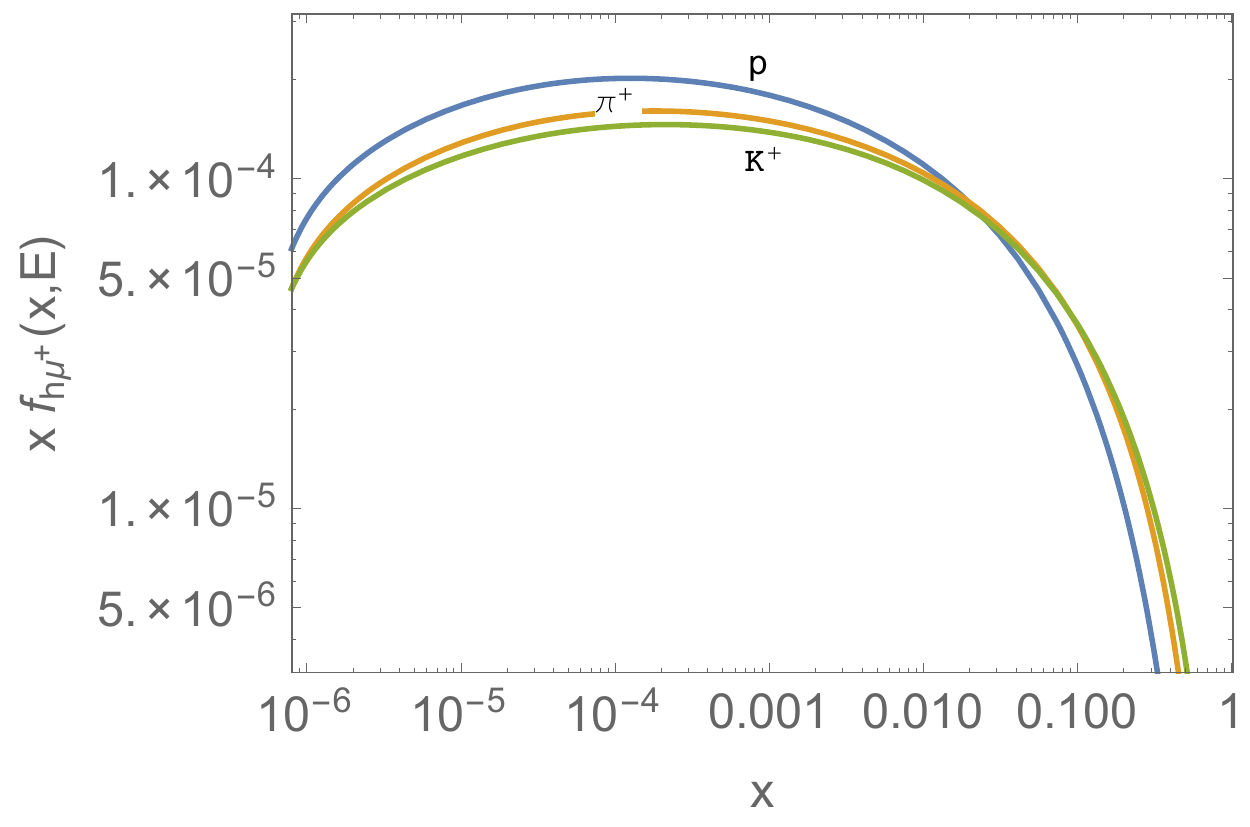}
\end{center}
\vspace{-0.5cm}
\caption{Unflavored meson yield in proton-air 
collisions at $10^6$ GeV obtained with EPOS-LHC (left), muon yield in
their decays (right), and muon yield through unflavored 
mesons in proton, pion and kaon collisions with air at $10^6$ GeV (lower).
}
\label{fig3}
\end{figure}

In Fig.~3 we plot the yields of unflavored mesons in proton-air collisions at 
$10^6$ GeV together with 
their decay yields into muons \cite{Illana:2010gh}. Since they decay almost instantly, we
can obtain the muon yield in hadron collisions through unflavored mesons as
\beq
f_{h\mu}(x,E) = \int_x^1 {\rm d}x'\; {1\over x'}\;\sum_{h'=\eta,\rho,...}
f_{hh'}(x',E/ x)\; f_{h'\mu}(x/x')\,.
\label{conv}
\eeq
We include for illustration also in Fig.~3 the muon yield in proton, pion and kaon collisions with 
air at $10^6$ GeV.

\noindent {\it (iv) Photon conversion into a muon pair.} \\
High energy photons appear in hadron collisions mostly 
through $\pi^0$ and $\eta$ decays. We obtain with EPOS-LHC that photons take
20.6\% of the energy in proton-air collisions at $10^6$ GeV, and that this
percentage is even larger in pion and kaon collisions (24.4\% and 24.1\%, respectively).
The gamma conversion length into $e^+e^-$ pairs 
is then 
\beq
{1\over \lambda_\gamma^{ee}(E,\rho)} = {7/9-b/3\over X_0} \; S^{\rm tot}_\gamma(E,\rho)
\eeq
where $S^{\rm tot}_\gamma(E,\rho)$ is the LPM suppression 
\cite{Stanev:1982au,Gerhardt:2010bj} (see next
section), $X_0=37.1$ g/cm$^2$
and (in air) $b=0.012$. The 
conversion into muon pairs will appear suppressed by a factor of
$m_e^2/m_\mu^2$, $\lambda_\gamma^{\mu\mu}\approx 2\times 10^6$ g/cm$^2$, 
and the fraction of energy going to each muon is distributed \cite{Rossi:1941zza}
\beq
f_{\gamma \mu^+}(x,E)=
{\lambda_{\gamma}^{\rm int}(E,\rho)\over X_0}\, \left({m_e\over m_\mu}\right)^2 
\left({2\over 3}-{b\over 2} + \left( {4\over 3} + 2 b \right)
\left( x- {1\over 2}\right)^{\!\!2} \,\right)\,.
\eeq
This is then a very rare process ($m_e^2/m_\mu^2=2.4\times 10^{-5}$) but one
where all the photon energy goes into muons. Notice also that the LPM effect
will favor the conversion into
muons (relative to electrons) at energies above
\beq
E_{\rm LPM}\approx 7.7 \;{X_0\over \rho}\;\; {\rm TeV/cm.}
\eeq

\noindent {\it (v) Photon conversion into a vector meson decaying into muons.} \\
In the previous process the photon fluctuates into a virtual muon pair that goes
on shell after an EM interaction with an air nucleus. However, the photon may also 
fluctuate into a $q \bar q$ pair, {\it i.e.}, a virtual vector meson that becomes real
after a hadronic (pomeron mediated)
interaction with the nucleus. This fluctuation is less likely 
due to the larger mass of the meson, but the suppression is partially
compensated by the larger coupling in the hadronic process.

It turns out that in 1 out of 400 collisions (and even more often at low energies
near a hadronic resonance) 
the photon behaves like a rho meson. The most frequent  photonuclear 
collision is then an inelastic
process resulting into a multiplicity of pions, but over $10\%$ of them 
are exclusive ($\gamma p\to \rho p$) or dissociative ($\gamma p\to \rho X$) 
conversions where the $\rho$ meson gets almost all of the photon energy. 
We will also consider the $\gamma$ 
conversion into a $J/\psi$ meson (a $c\bar c$ state), more rare than the
$\rho$ (specially at lower energies)
but with a much larger branching ratio into $\mu^+\mu^-$ (around a $5.9\%$).

In our estimate for these processes
we have extrapolated  the HERA observations at
$\sqrt{s}< 300$ GeV \cite{Merkel:1999hd,Aktas:2005xu} up to  $\sqrt{s}< 300$ TeV
using a two-pomeron scheme:
\beqa
\sigma_{tot}(\gamma p)&=&69.0 \,s^{0.08}+175 \,s^{-0.60},\cr
\sigma(\gamma p\to \rho p)&=&4.9\,s^{0.11}+21 \,s^{-0.40},\cr
\sigma(\gamma p\to J/\psi\, p)&=&0.0016\,s^{0.41}\,,
\eeqa
where $s$ is given in GeV$^2$ and the cross sections in $\mu$b. To include
dissociative conversions \cite{Chekanov:2002rm,Bendova:2018bbb} 
we have just added a $60\%$ to the exclusive
cross sections above, and we have assumed that the scaling to go from a proton 
to a nucleus target coincides with the one in pion collisions. 

\section{Cascade equations and muon flux}
We will solve numerically the (longitudinal) cascade equations for 
15 hadron species 
$h$ ($p$, $n$, $\bar p$, $\bar n$,  $\pi^\pm$,  $K^\pm$, $K_L$, $D^\pm$,  
$D^0$, $\bar  D^0$, $D_s^\pm$,
 $\Lambda_c^\pm$), photons, electrons and muons, with $\mu^\pm$
 from the prompt decay of unflavored mesons included in the yields
$f_{h\mu}(x,E)$ and  $f_{\gamma \mu}(x,E)$ (see previous section).
The initial flux ($t=0$) will correspond to a single primary of energy $E_0$,
whereas the secondary fluxes will be defined for $E\le E_0$.
The generic equations are \cite{Lipari:1993hd}
\beqa
{{\rm d}\Phi_i(E,t)\over {\rm d} t} &=&-\,{\Phi_i(E,t)\over 
\lambda_i^{\rm int}(E)} -
{\Phi_i(E,t)\over \lambda_i^{\rm dec}(E,t)} +
\sum_{j=h,\gamma,e} \int_{E/E_0}^1 {\rm d}x\; {f_{ji}(x,E/x)\over x}\;
{\Phi_{j}(E/x,t)\over \lambda_{j}^{\rm int}(E/x)} +\nonumber\\
&&\sum_{k=h} \int_{E/E_0}^1 {\rm d}x\; {f_{ki}^{\rm dec}(x,E/x)\over x}\;
{\Phi_{k}(E/x,t)\over \lambda_{j}^{\rm dec}(E/x,t)}\,,
\eeqa
where $\Phi_i={\rm d} N_i/dE$, $t$ is the slant depth and the
interaction/decay lengths are expressed in g/cm$^2$. We will focus on
particles with energy between 1 TeV and the energy of the primary. 

\begin{figure}[!t]
\begin{center}
\includegraphics[width=0.5\linewidth]{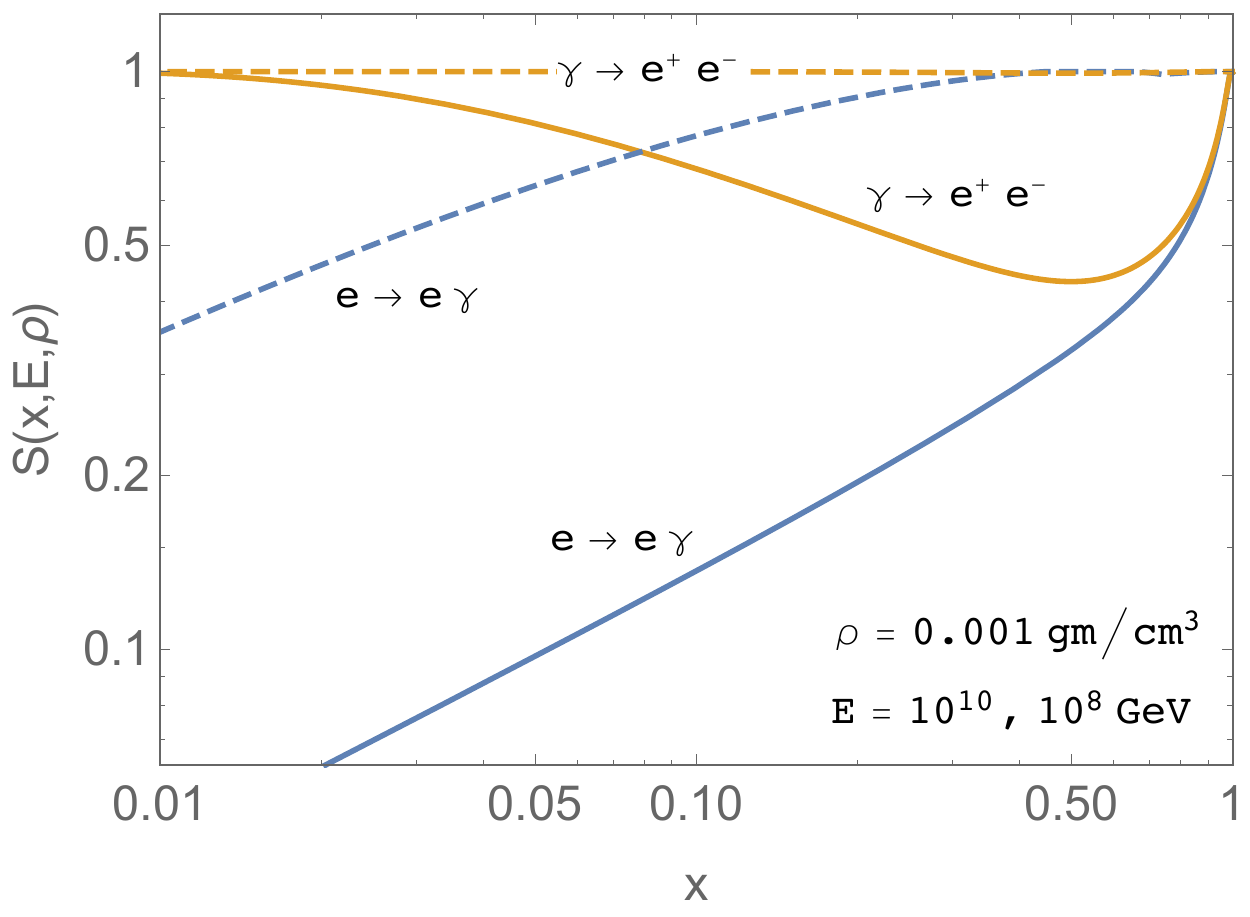}
\end{center}
\vspace{-0.5cm}
\caption{LPM reduction in ${\rm d} \sigma(e\to e\gamma)/{\rm d}  x$ and 
${\rm d}  \sigma(\gamma\to e^+e^-)/{\rm d}  x$ in air at 
$\rho=0.001$ g/cm$^3$ for $E=10^{10}$ GeV (solid), $10^{8}$ GeV (dashes), 
with $x$ the fraction
of energy carried by the final $\gamma$ and $e^-$, respectively. We have
used the (non-recursive) expressions in \cite{Stanev:1982au}.
}
\label{fig4}
\end{figure}

The EM component of the shower will be started in hadronic collisions 
through the decay of neutral mesons and other hadronic resonances
(we obtain $f_{h\gamma}(x,E)$ from a fit to EPOS-LHC simulations);
we neglect the production of electrons in 
hadron and muon decays and also the photonuclear collisions of electrons. 
The cascade equations for photons and electrons read then \cite{Gaisser:1990vg}
\beqa
{{\rm d}\Phi_\gamma(E,t)\over {\rm d} t} =-\,{\Phi_\gamma(E,t)\over 
\lambda_\gamma^{\rm int}(E,t)}  \! \!&+&\! \!
\sum_{j=h} \int_{E/E_0}^1 {\rm d}x\; {f_{j\gamma}(x,E/x)\over x}\;
{\Phi_{j}(E/x,t)\over \lambda_{j}^{\rm int}(E/x)} \nonumber\\
\! \!&+&\! \! \int_{E/E_0}^1 {\rm d}x\; {f_{e\gamma}(x,E/x,t)\over x}\;
{\Phi_{e}(E/x,t)\over \lambda_{e}^{\rm int}(E/x,t)}\,,
\eeqa
and
\beqa
{{\rm d}\Phi_e(E,t)\over {\rm d} t} =-\,{\Phi_e(E,t)\over 
\lambda_e^{\rm int}(E,t)}  \! \!&+&\! \!
 \int_{E/E_0}^1 {\rm d}x\; {2\,f_{\gamma e}(x,E/x,t)\over x}\;
{\Phi_{\gamma}(E/x,t)\over \lambda_{\gamma}^{\rm int}(E/x,t)} \nonumber\\
\! \!&+&\! \! \int_{E/E_0}^{1-x_{\rm min}} {\rm d}x\; {f_{e\gamma}(1-x,E/x,t)\over x}\;
{\Phi_{e}(E/x,t)\over \lambda_{e}^{\rm int}(E/x,t)}\,,
\eeqa
where $x_{\rm min}(E)={E_{\rm min}^\gamma/ E}$ and 
we have used that $f_{ee}(x,E)=f_{e\gamma}(1-x,E)$. 
The interaction lengths are 
\beqa
&&{1\over  \lambda_{\gamma}^{\rm int}(E,t)}= {7-3b\over 9X_0} \left( S^{\rm tot}_\gamma(E,\rho)+
{m_e^2\over m_\mu^2} \right)+
{\sigma^{\rm had}_{\gamma A}\over m_A};\nonumber \\
&&{1\over  \lambda_{e}^{\rm int}(E,t)}=  {\int_{x_{\rm min}}^1\!{\rm d}x\;\phi(x)\over X_0}
\,S_e^{\rm tot}(E,\rho),
\eeqa
with $m_A$ the target mass (in grams) in an average hadronic collisions in the air
({\it i.e.}, $A=14.6$), while the EM yields are \cite{Rossi:1941zza}
\beqa
&&f_{\gamma e}(x,E,t)= {\lambda_{\gamma}^{\rm int}(E,t)\over X_0} \,
S_\gamma(x,E,\rho) \,\psi(x);\hspace{0.7cm}\psi(x)=
{2\over 3}-{b\over 2} + \left( {4\over 3} + 2 b \right)
\left( x- {1\over 2}\right)^{\!\!2},\nonumber \\
&&f_{e \gamma}(x,E,t) =   {\lambda_{e}^{\rm int}(E,t)\over X_0}\,
S_e(x,E,\rho) \,\phi(x)
;\hspace{0.7cm}\phi(x)=x+{1-x\over x}\,\left({4\over 3}+2 b\right).
\eeqa
The factors $S^{\rm tot}_{\gamma,e}$ and $S_{\gamma,e}$ above
express the LPM reduction in the total and the differential cross sections 
for $\gamma \to e^+e^-$ and $e \to e\gamma$ in the air. In Fig.~\ref{fig4}
we plot $S_{\gamma,e}(x,E,\rho)$ for $E=10^8,10^{10}$ GeV and
$\rho=0.001$ g/cm$^3$; the suppression in the total cross section is
in those cases 
$S^{\rm tot}_{e}=0.043$ and  $S^{\rm tot}_{\gamma}=0.58$ at $10^{10}$ GeV but
just $S^{\rm tot}_{e}=0.33$ and  $S^{\rm tot}_{\gamma}=0.99$ at $10^{8}$ GeV.

We obtain the nuclear cross sections needed in the hadronic 
interaction lengths ($\lambda_h^{\rm int}=m_A/\sigma_{hA}$)
with EPOS-LHC, and we use isospin symmetry to deduce the whole set of yields from 
the ones in $p$, $\pi^+$ and $K^+$ collisions. The charmed hadron yields have been
deduced with SIBYLL2.3C (we have normalized the EPOS-LHC yields of light mesons
and baryons to subtract
the energy taken by these $D$ mesons and $\Lambda_c$ baryons).
As for the muons, we 
neglect energy loss as they propagate, but in the next section we will calculate the
probability for a catastrophic energy deposition in the air near the surface. 
For the atmosphere we assume \cite{Maeda:1973nz} (in g/cm$^3$) 
\beq
\rho(h) =
\left\{\begin{array}{ll}
1.210\times 10^{-10} \left(44.33-h\right)^{4.253}, &  h <11 \;{\rm km;}\\
\\
2.053\times 10^{-3} \exp\left(-{h\over 6.344\;{\rm km}}\right),&  h >11 \;{\rm km}
\,.
\end{array}\right.
\eeq

We have taken 200 logarithmic bins of energy with $E_{\rm min}=10$ TeV
and 2500 linear bins of altitude 
with $h_0=70$ km, and we 
have checked that the transport through the atmosphere 
conserves the total energy in the shower. Our results are summarized in 
\begin{figure}[!t]
\begin{center}
\includegraphics[width=0.471\linewidth]{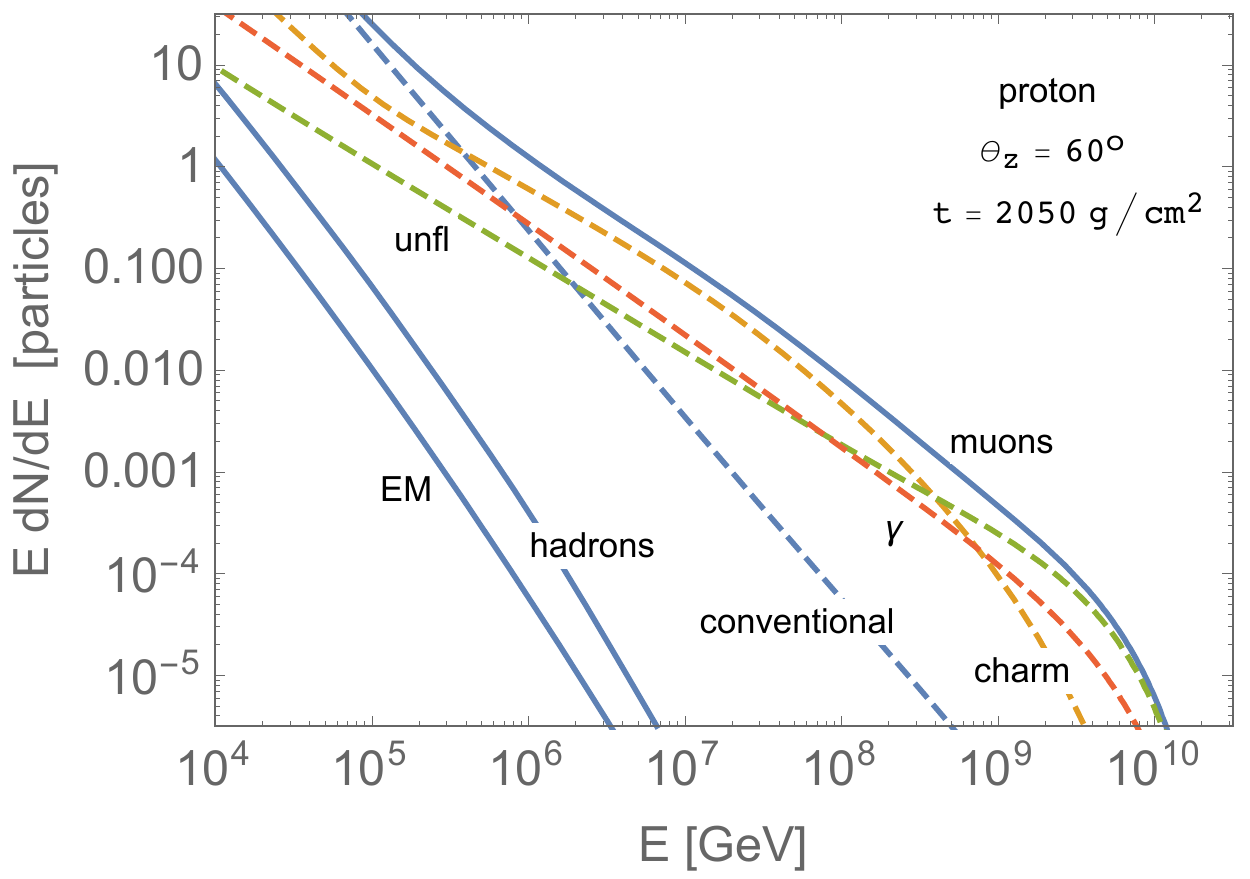}\hspace{0.8cm}
\includegraphics[width=0.471\linewidth]{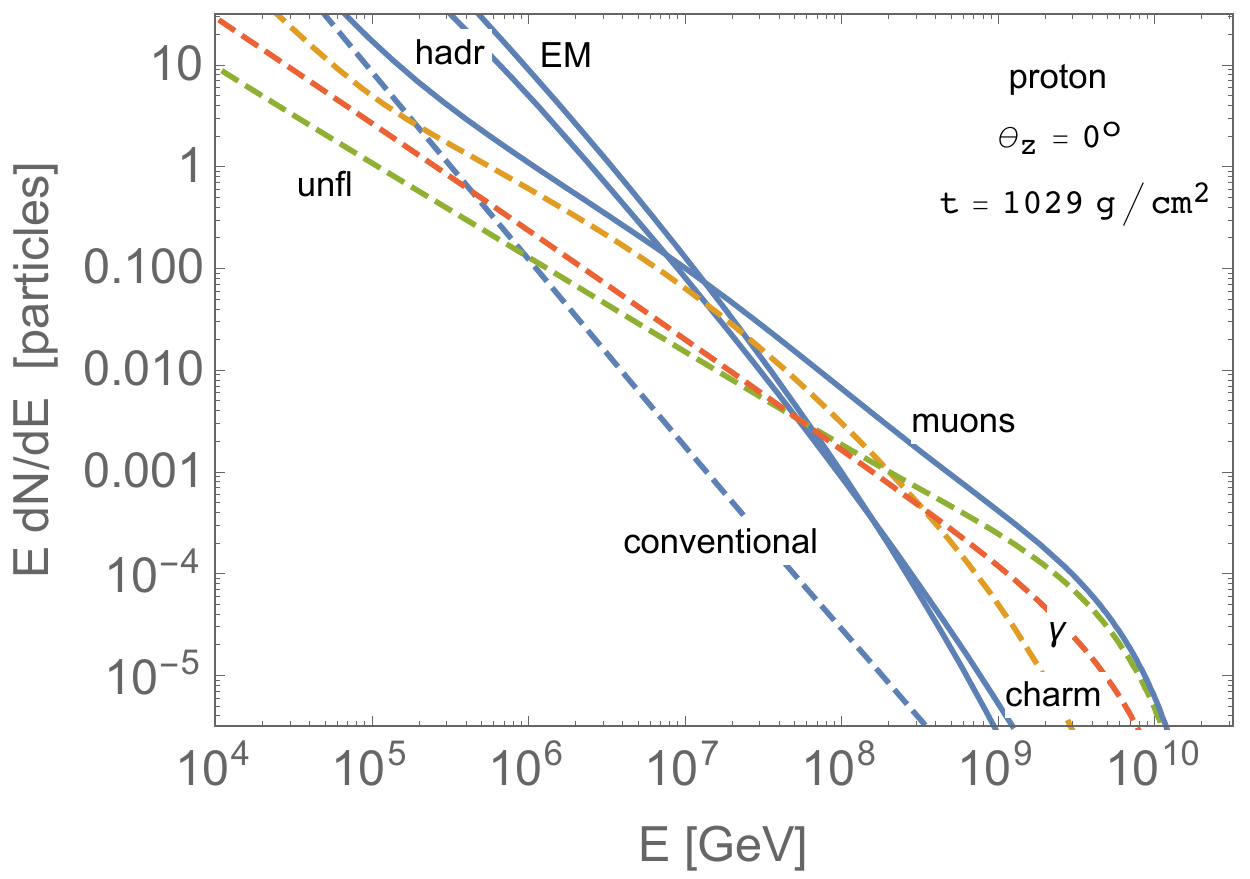}
\includegraphics[width=0.471\linewidth]{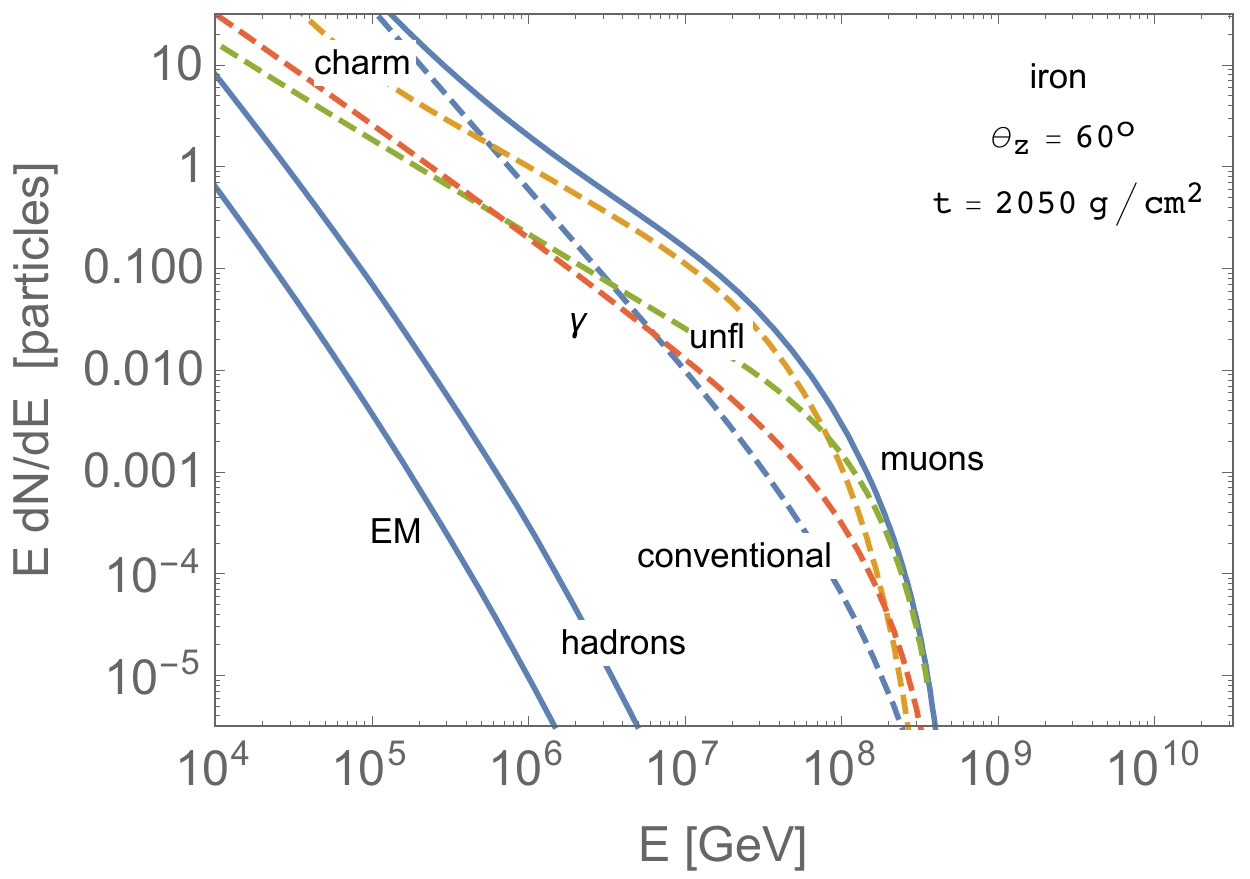}\hspace{0.8cm}
\includegraphics[width=0.471\linewidth]{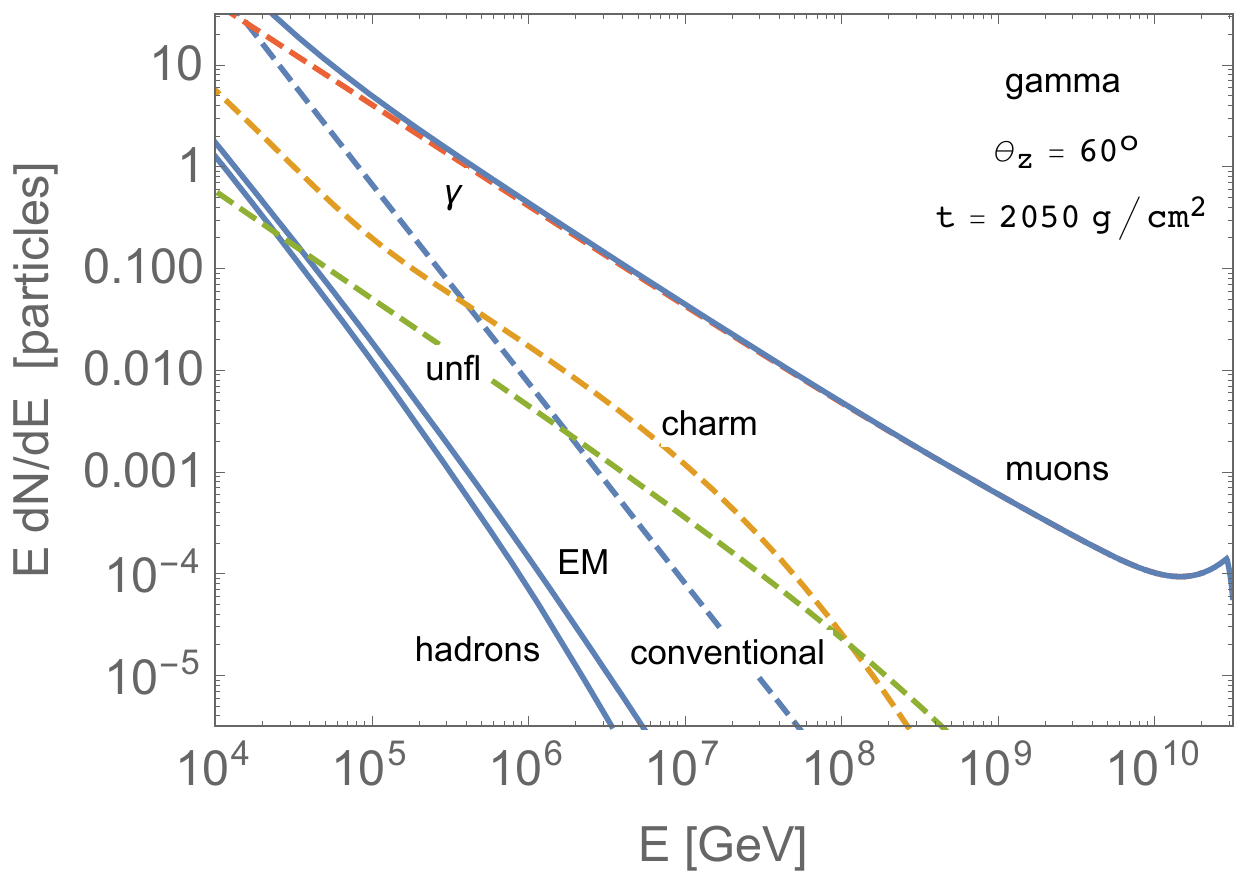}
\end{center}
\vspace{-0.5cm}
\caption{Particle count at  the ground level for a $10^{10.5}$ GeV
proton (top left), iron nucleus (bottom left) or gamma ray (bottom right)
coming from $\theta_z=60^\circ$, and for a vertical 
proton  of the same energy (top right).
The EM line includes photons and electrons, and 
the dashed lines indicate the 
different contributions to the muon flux (the $\gamma$ line includes the muons from
$\gamma  \to \mu\mu$ and from $\gamma\to \rho, J/\psi \to \mu\mu$). 
}
\label{fig5}
\end{figure}
Fig.~\ref{fig5}. There we plot the particle flux (number of particles per unit energy)  
at the ground level for several primaries, all of them with $E=10^{10.5}$ GeV. 

The upper figures include 
a proton from zenith angles $\theta_z=60^\circ$ (left)
and $\theta_z=0^\circ$ (right). We find a total of 0.0065 muons with
$E\ge 10^8$ GeV in the first case and 0.0052 muons in a vertical shower. 
This implies, respectively, 
that around 1 in 150 or 1 in 190 showers include such a muon. The contribution to
this muon count from unflavored mesons and gamma conversions is basically
independent from the shower inclination, while the charm contribution has a 40\%
reduction for vertical showers (it goes from 0.0030 to 0.0018).  In the inclined proton
shower charm decays generate 46\% of the $E>10^8$ GeV muons, 
unflavored decays 30\% and photon conversions 23\%, being the
contribution from $\gamma\to \mu^+\mu^-$  three times larger 
than the one from $\gamma\to J/\psi \to  \mu^+\mu^-$.

In the lower figures we plot the fluxes for 
iron (left) or photon (right) showers of also $E=10^{10.5}$ GeV, 
both from a zenith inclination $\theta_z=60^\circ$. The number of 
muons with energy $E>10^8$ GeV is, respectively, 0.0010 and 0.0053. 
This means that
only 1 in 1000 iron showers or 1 in 186 photon showers
include such a muon. In an iron primary charm decays contribute
a 28\% to this muon count, whereas in gamma showers 99\% of the muons
come from gamma conversions (71\% in EM interactions and 28\% through $J/\psi$
decays).
We also see that the conventional contribution from pion and kaon decays 
is negligible both for proton or gamma primaries,
but  it is  more significant (2.1\%) in iron showers. We find remarkable that
in photon showers the conversions $\gamma\to \mu^+\mu^-$  and 
$\gamma\to J/\psi \to  \mu^+\mu^-$ give many more muons 
than pion and kaon decays even at lower energies 
(see the discussion on muons from 
inelastic photonuclear collisions in \cite{Cornet:2015qda}).
We notice as well that the LPM effect on the muon count is negligible in
iron or proton showers, while in the average
$10^{10.5}$ GeV photon shower it increases the number of 
these very high energy muons in just a $2\%$.

\section{Muon energy depositions near the surface}
\begin{figure}[!t]
\begin{center}
\includegraphics[width=0.5\linewidth]{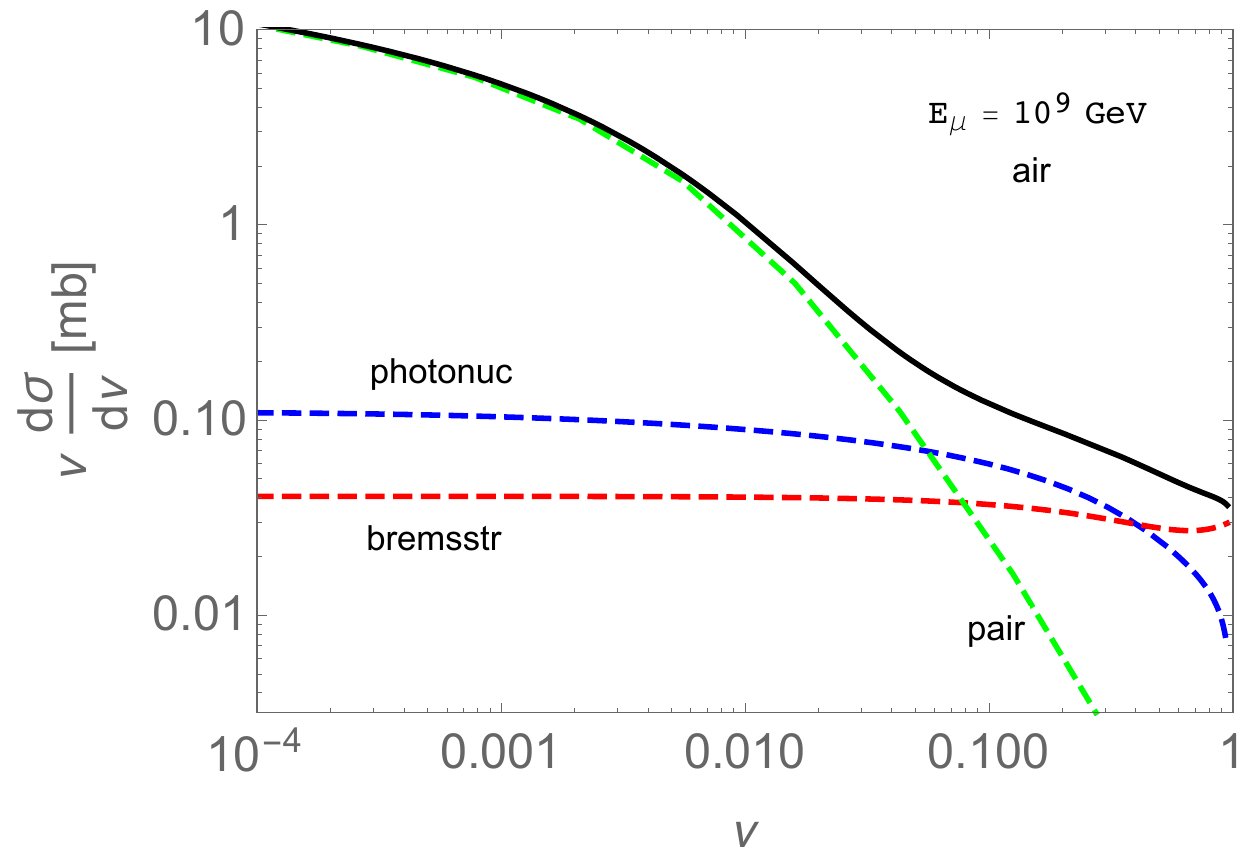}
\end{center}
\vspace{-0.5cm}
\caption{Cross section to deposit a fraction $\nu$ of energy in muon
collisions in the air. 
}
\label{fig6}
\end{figure}

High-energy muons may radiate a significant fraction of their energy through
three different processes: bremsstrahlung, pair production or photonuclear 
interactions \cite{Groom:2001kq}. The first two processes would start an EM shower, whereas
photonuclear collisions would define a hadronic sub-shower. In 
inclined events these energy 
depositions could occur very deep in the atmosphere, 
when most of the shower energy has been absorbed.
In Fig.~\ref{fig6} we plot the differential
cross section for these radiative processes, being $\nu$ the fraction of the muon
energy deposited in the air.

The probability that a muon of energy $E$ has an interaction within a 
depth $\Delta X=500$ g/cm$^2$ where it radiates
a minimum energy $E_0$ is
\beq
p(E_0,E)={\Delta X\over m_A} \int_{E_0/E}^1 {\rm d}\nu\; {{\rm d}\sigma\over {\rm d}\nu}\,,
\eeq
with values larger than 1 expressing the average number of depositions.
For an incident flux $\Phi_\mu(E)$ the probability to have the same type of
energy deposition 
is then
\beq
p(E_0)={\Delta X\over m_A} \int_{E_0}^\infty {\rm d}E\;\Phi_\mu(E)
\int_{E_0/E}^1 {\rm d}\nu\; {{\rm d}\sigma\over {\rm d}\nu}\,.
\eeq
Our results for the 3 primaries considered in the previous section are the following.
In a $10^{10.5}$ GeV proton shower from $\theta_z=60^\circ$ 
the probability to have an energy deposition above $10^6$ GeV is 0.17,
{\it i.e.}, we can expect such an anomaly in one 
out of 6 proton showers. $E\ge 10^7$ depositions would be much less frequent:
around in 1 out of 330 showers. In contrast, 
the $E\ge 10^6$ GeV depositions would occur in only 1 out of 92 iron
showers, and the deposition would be above $10^7$ GeV in one per 7000 iron
showers. 
For a photon primary the spectrum of high-energy 
muons is much harder than in hadron showers (see Fig.~4). Since a 
$10^9$--$10^{10}$ GeV
muon loses around 0.4\% of its energy per 1000 g/cm$^2$ of air,
{\it most} muons in this energy range (appearing in around 1 in 200 photon 
showers) 
will start {\cal O}$(10^7\;{\rm GeV})$ EM mini-showers near the ground.

\section{Cascade equations versus Monte Carlo simulations}
We would like to briefly address some of the limitations and the 
validity of the method that we have used for the 
study of EASs. Our simplified cascade equations 
are a fast and flexible way to estimate the relative relevance of a given
effect, but they can not substitute the more precise results obtained
with Monte Carlo codes like AIRES \cite{AIRES} and CORSIKA \cite{Heck:1998vt} or with 
hybrid models (combining Monte Carlo methods with 
cascade equations) 
like CONEX \cite{Bergmann:2006yz}
and SENECA \cite{Drescher:2002cr}.
Our method seems specially useful to estimate the relative effect
of a rare process ({\it e.g.}, photon conversions 
into muon pairs) whose accurate study with simulations
would require a very large statistics.

We have used 1-dimensional cascade equations that neglect the lateral versus
the longitudinal development of the shower. As a consequence, we obtain 
a poorer approximation at lower energies, where the transverse
momentum of the particles may be relatively important and imply a 
larger lateral displacement. 
In addition, to simplify
the equations we have not included effects, 
like energy loss by ionization, that are also important at low
energies. The 
energy binning, the finite value of the depth intervals or 
the simple model that we have used for the 
atmosphere are sources of uncertainty as well.
\begin{figure}[!t]
\begin{center}
\includegraphics[width=0.44\linewidth]{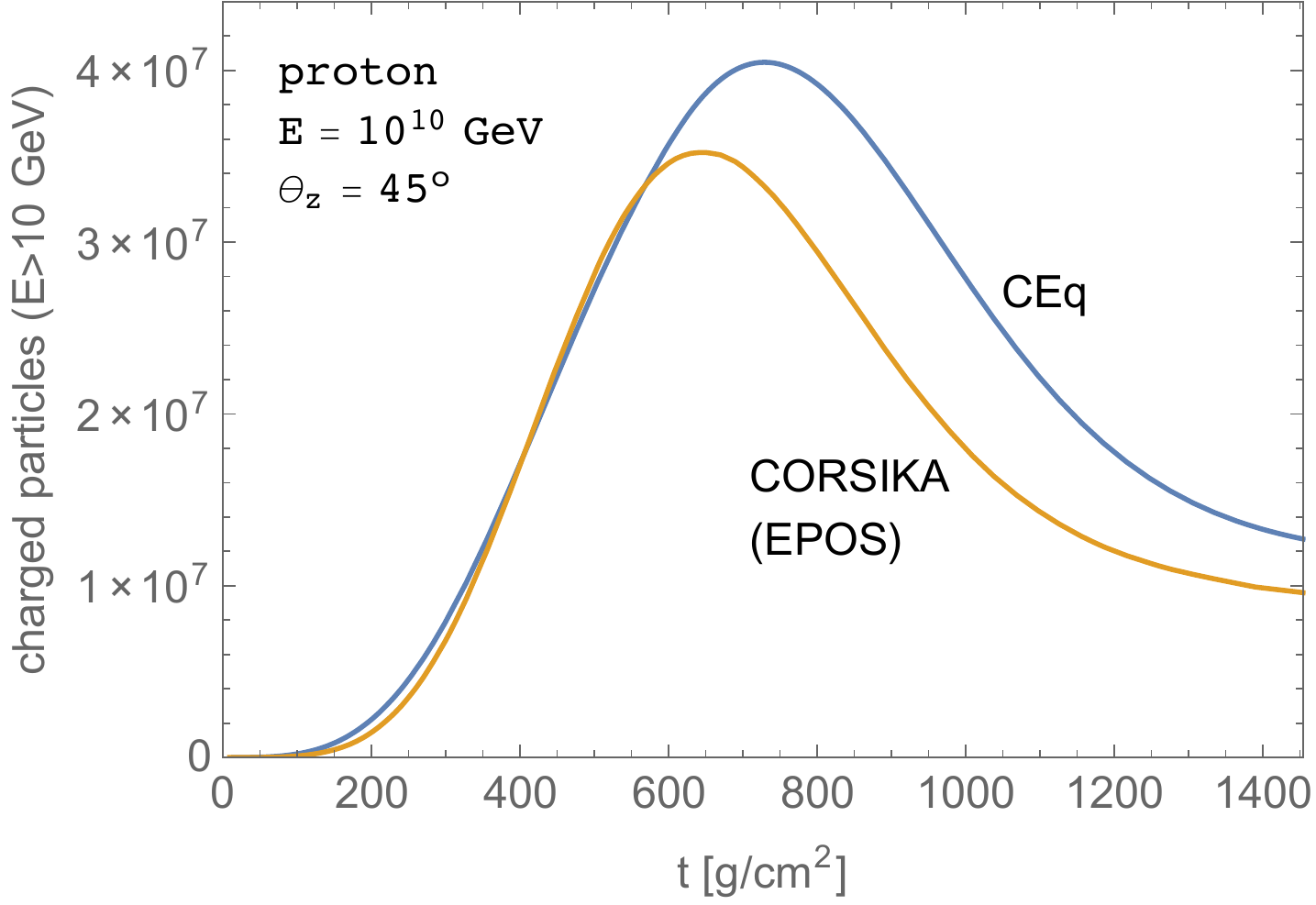}
\hspace{0.8cm}\includegraphics[width=0.45\linewidth]{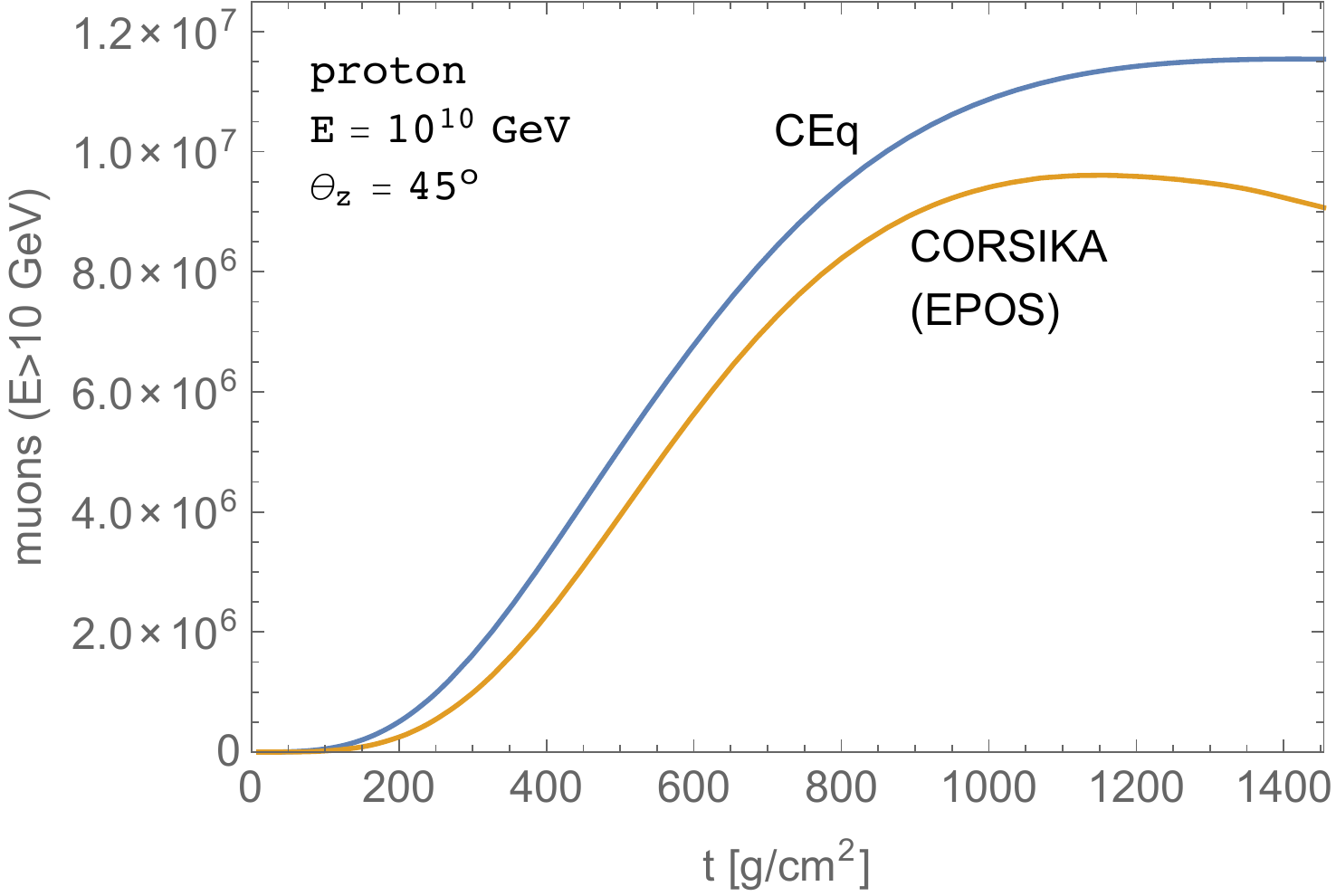}
\end{center}
\vspace{-0.5cm}
\caption{Number of charged particles (left) or muons (right)
with $E>10$ GeV in a $10^{10}$ GeV proton
shower at different slant depths obtained with
cascade equations and CORSIKA.
}
\label{fig7}
\end{figure}

To calibrate the accuracy of our results we have solved the cascade 
equations for a $10^{10}$ GeV proton shower from $\theta_z=45^\circ$ and
have extended the range of energies down to $10$ GeV. We have then
averaged 30 showers simulated with CORSIKA 
(running with the EPOS-LHC option and a $10^{-6}$ thinning) 
with the same primary and minimum energy. In  Fig.~\ref{fig7} 
we plot the total number of charged particles (including muons) with
$E>10$ GeV at different atmospheric depths (left), together with the evolution
in the number of muons (right).
The Monte Carlo
simulations imply a value of 
$X_{\rm max}$ and a number of particles 20\% smaller than the one obtained
with cascade equations. 

\begin{figure}[!t]
\begin{center}
\includegraphics[width=0.45\linewidth]{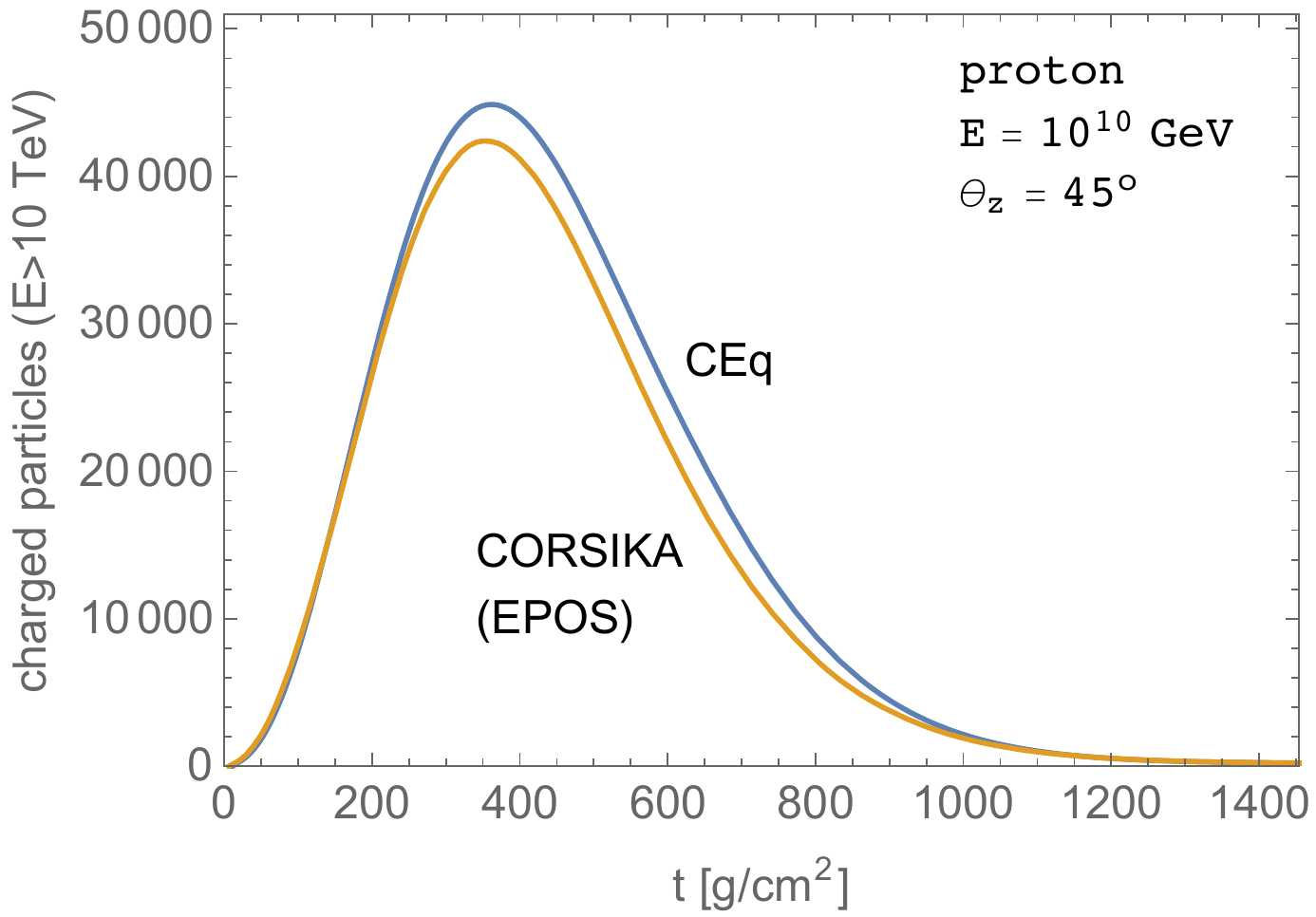}
\hspace{0.8cm}\includegraphics[width=0.43\linewidth]{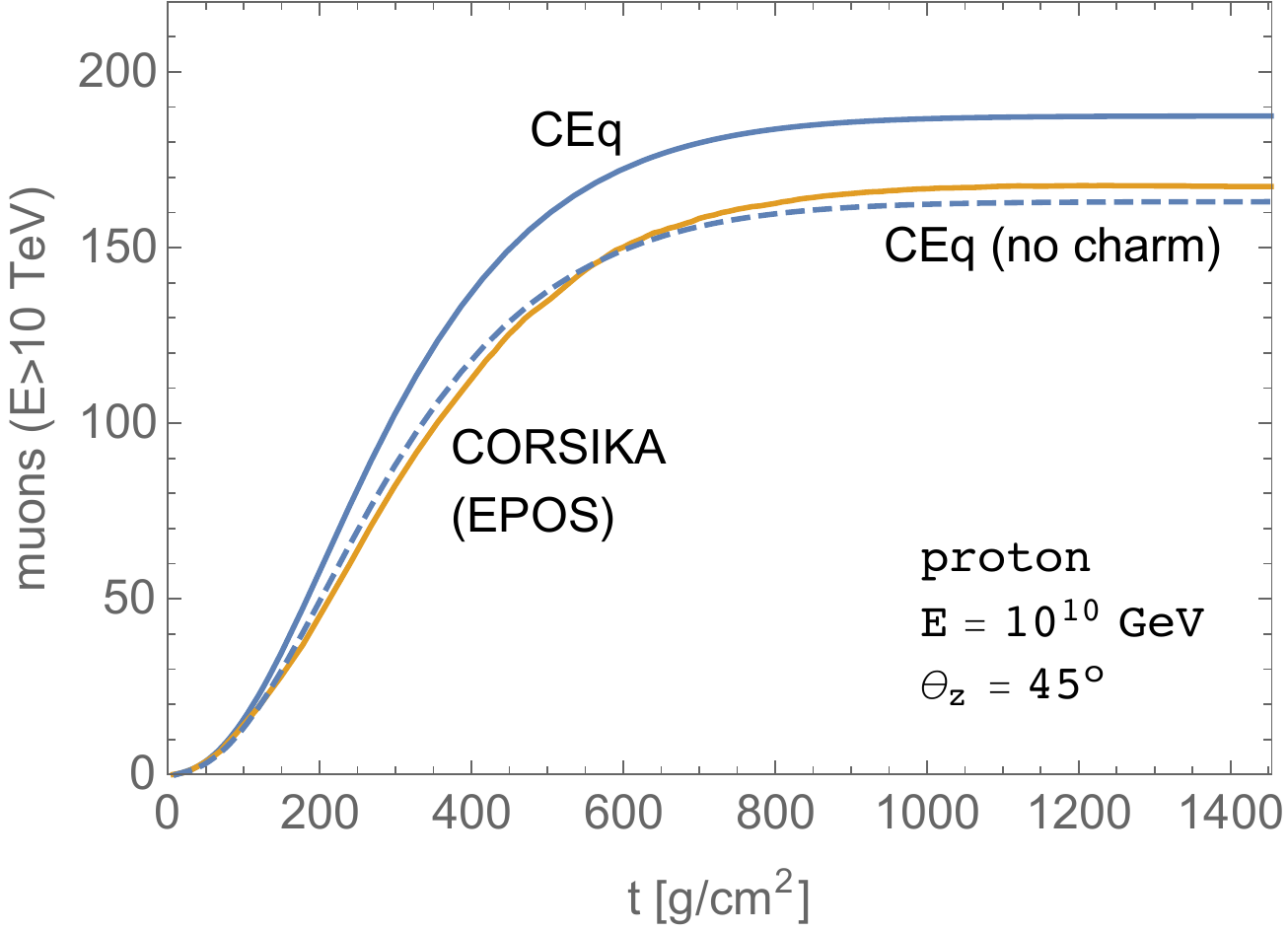}
\end{center}
\vspace{-0.5cm}
\caption{Number of charged particles (left) or muons (right) 
with $E>10$ TeV in a $10^{10}$ GeV proton
shower at different slant depths obtained with
cascade equations and CORSIKA. In dashes, we have subtracted
the muons from charm decays.
}
\label{fig8}
\end{figure}

The analysis that we have presented in the previous sections, however,
involves muons of much higher energy,
where we expect more accuracy. To confirm that we have run 30 more
proton showers but taking a larger value of the minimum energy: $E>10$ TeV.
The comparison with the results from the cascade equations, in Fig.~\ref{fig8},
show now a much better agreement. 
We obtain that the differences in $X_{\rm max}$ and in number of charged 
particles are within a $5\%$. The difference is more significant in the
number of muons with $E>10$ TeV, but this is due to the absence of charm in
the CORSIKA simulation with the EPOS-LHC option; 
once we subtract the muons from charm (yields deduced with
SIBYLL 2.3C) we obtain the
almost perfect agreement shown in Fig.~\ref{fig8}-right.

\section{Summary and discussion}
In studies of EASs
the connection between the primary CR and the secondary particles 
observable at 
the surface usually relies on a Monte Carlo 
simulation.
It is then necessary to make sure that all the relevant effects are included,
and 1-dimensional cascade equations may be a useful tool. 
We find that to account for the most energetic muons in a shower
one has to include, in addition to 
charmed hadrons, the rare decays of unflavored mesons and the photon 
conversions into muon pairs and $J/\psi$ mesons. This is clear in photon
showers at all muon energies (the conversions 
produce more muons than pion and kaon decays even at $10^4$ GeV,  see 
Fig.~\ref{fig5}), but also in proton showers at $E\ge 10^9$ GeV. Remarkably, 
the EAS simulator CORSIKA \cite{Heck:1998vt}
with the SIBYLL 2.3C option would include all of
the processes discussed here except for the quasielastic conversion of
a photon into a $J/\psi$ meson. Notice, however,
that for proton or iron primaries this contribution is subleading: photon conversions account
for $23\%$ of all muons of $E>10^8$ GeV in the shower, 
and only $24\%$ of
those muons come from $J/\psi$ decays. Although all the muon sources are 
then standard and known, we think that the complete analysis of 
their (energy dependent) relative weight can not be found in previous
literature.

The fluctuations in the value of $X_{\rm max}$ in EASs are caused 
by the details (interaction point, multiplicity, inelasticity) in the first 
hadronic collisions high in the atmosphere. 
Once the shower has reached $X_{\rm max}$,
its hadron and EM components include millions of particles sharing most of 
the shower energy, and the individual fluctuations appear then {\it averaged}. 
Hadrons or electrons are unable to produce large fluctuations in any
observable after $X_{\rm max}$ because all of them have an energy 
much smaller than the total shower energy at that depth. What we discuss 
here is arguably the only possible source of fluctuations 
in the longitudinal development of an EAS {\it after} $X_{\rm max}$: very 
high energy muons with catastrophic depositions 
near the surface. Only muons can take a large amount of energy and 
deposit it near the ground in an inclined shower. This effect would 
make possible that two inclined showers with the same primary and the same 
value of $X_{\rm max}$ evolve different as 
they get to the ground. Our analysis here intends to quantify the frequency
of such events, and it may motivate a more complete study (using Monte Carlo or
hybrid models)
of its possible relevance at current or future EAS observatories.

Our analysis is based on a numerical solution to the
longitudinal cascade equations 
 through the atmosphere. At high energies the method is able to incorporate
easily  and precisely each one of the new effects (Monte Carlo 
methods may be less efficient to capture the rare effects discussed here). 
Our results are consistent with the
ones in \cite{Fedynitch:2018cbl}, 
that focus on the total (inclusive) atmospheric muon flux  and emphasize
the relevance of unflavored meson decays, although photon conversions
into muon pairs and $J/\psi$ mesons are not included there.

We estimate that 1 in 43 inclined events started by 
a $10^{10.5}$ GeV proton primary 
contains a muon taking more than $0.1\%$ of the total shower energy 
({\it i.e.}, $E>10^{7.5}$ GeV), that 1 in 6 proton
showers include a radiative energy deposition above 
 $10^6$ GeV within 
500 g/cm$^2$ near the surface, and that in 1 in 330 showers this 
deposition is above $10^7$ GeV. These frequencies are different for iron or
gamma primaries (only 1 in 60 iron showers or 61 photon showers
includes such a muon). The appearance of an 
EM shower after most of the parent EAS
has been absorbed ({\it i.e.}, beyond 1500 g/cm$^2$) 
could introduce rare fluctuations in the muon to electron count at 
the ground level \cite{Canal:2016bvv},
something that may be measured with enough accuracy after the upgrade in the surface
detectors at AUGER \cite{ThePierreAuger:2015rma}. Since no hadrons
can keep $10^6$--$10^7$ GeV after such depth, the fluctuations may provide an
indirect signal of the hadronic processes 
discussed here. In particular, they may
be an interesting channel to search for (the so far elusive) atmospheric charm.
These very energetic muons are also of interest 
at $\nu$ telescopes, where a determination
of the  $E\ge 100$ TeV muon flux  and its correlation with the
neutrino flux at the same energies could also reveal a contribution from  
atmospheric charm.

Hadronic simulators like EPOS-LHC or SIBYLL give
predictions at extreme energies, well beyond the ones achieved
at particle colliders. The muons discussed here are a probe of 
those energies. We think that this could make them useful in the 
study of both the hadronic cross sections in this regime 
and  the 
composition of the highest energy CRs.

\section*{Acknowledgments}
The authors woud like to thank T.~Pierog and R.~Ulrich for providing 
the {\it Cosmic Ray Monte Carlo} package \cite{crmc}.
This work has been supported by MICINN of Spain 
(FPA2016-78220, RED2018-102661-T) 
and by Junta de Andaluc\'\i a (SOMM17/6104/UGR and FQM101). 
CG and JSM acknowledge a grant from {\it Programa Operativo de Empleo Juvenil}
(Junta de Andaluc\'\i a) and MG an 
{\it Iniciaci\'on a la Investigaci\'on} fellowship from the University of Granada.

\end{document}